\documentclass[aip,
 amsmath,amssymb,
 reprint,%
]{revtex4-1}
\usepackage{natbib}       % 参考文献宏包
\usepackage{url}          % 处理 URL 特殊字符
\usepackage{hyperref}     % 可选，生成超链接（和 url 二选一即可，hyperref 包含 url）
\usepackage{graphicx}% Include figure files
\usepackage{dcolumn}% Align table columns on decimal point
\usepackage{bm}% bold math
\usepackage{color}
\usepackage[utf8]{inputenc}
\usepackage[T1]{fontenc}
\usepackage{mathptmx}
\usepackage{etoolbox}

\makeatletter
\def\@email#1#2{%
 \endgroup
 \patchcmd{\titleblock@produce}
  {\frontmatter@RRAPformat}
  {\frontmatter@RRAPformat{\produce@RRAP{*#1\href{mailto:#2}{#2}}}\frontmatter@RRAPformat}
  {}{}
}%
\makeatother
\begin{document}

\preprint{AIP/123-QED}

\title{Phason-Driven Diversity of Nucleation Pathways in Icosahedral Quasicrystals}
% Force line breaks with \\
\author{Gang Cui}
\affiliation{Hunan Key Laboratory for Computation and Simulation in Science and Engineering, Key Laboratory of Intelligent Computing and Information Processing of Ministry of Education, School of Mathematics and Computational Science, Xiangtan University, Xiangtan, Hunan, China, 411105.}%Lines break automatically or can be forced with \\

\author{Lei Zhang}%
%\email{zhangl@math.pku.edu.cn}
\affiliation{Beijing International Center
for Mathematical Research, Peking University, Beijing, 100871, China.
}%

\author{Pingwen Zhang}%
\email{pzhang@pku.edu.cn}
\affiliation{Institute for Math \& AI, Wuhan, Wuhan University, Wuhan, Hubei, 430072, China.}%
\affiliation{School of Mathematical Sciences, Peking University, Beijing, 100871, China.}

\author{An-Chang Shi}%
\email{shi@mcmaster.ca}
\affiliation{Department of Physics and Astronomy, McMaster University, Hamilton, Canada, L8S 4M1.
}%

\author{Kai Jiang}
\email{kaijiang@xtu.edu.cn}
\affiliation{Hunan Key Laboratory for Computation and Simulation in Science and Engineering, Key Laboratory of Intelligent Computing and Information Processing of Ministry of Education, School of Mathematics and Computational Science, Xiangtan University, Xiangtan, Hunan, China, 411105.}%Lines break automatically or can be forced with \\

\date{\today}% It is always \today, today,

\begin{abstract}
The nucleation of quasicrystals remains a fundamental puzzle, primarily due to the absence of a periodic translational template. Here, we demonstrate that phasons—hidden degrees of freedom unique to quasiperiodic order—drive diverse nucleation pathways in icosahedral quasicrystals (IQCs). Combining a Landau free-energy model with the spring pair method, we compute distinct critical nuclei and their corresponding minimum energy paths. At low temperatures, a direct, symmetry-preserving pathway dominates. In contrast, higher temperatures promote a "symmetry detour" that reduces the nucleation barrier via a lower-symmetry critical nucleus. Remarkably, while the resulting bulk IQCs exhibit distinct real-space symmetries, they remain thermodynamically degenerate with identical diffraction patterns. We resolve this paradox within the high-dimensional projection framework, showing that phason shifts modulate real-space symmetry without altering bulk thermodynamics. Our findings establish phasons as the structural origin of pathway diversity, offering a new physical picture for the emergence of quasiperiodic order.
\end{abstract}

\maketitle
\section{Introduction}
Nucleation is the pivotal event in constructing the ordered  world, governing the formation of crystals, polymers, and proteins alike.~\cite{Karthika2016Review,Paul2017classical,WU2022Nucleation}
While nucleation in conventional  periodic crystals has been extensively studied~\cite{Sosso2016Crystal,James2019How,Du2024NonClassical}, the nucleation mechanisms of quasicrystals (QCs) remain far less understood.
QCs challenge the traditional paradigm of structural order, as they exhibit long-range order in the absence of translational periodicity~\cite{Shechtman1984}.
This aperiodic order originates from the projection of a higher-dimensional periodic lattice\cite{Walter2009Crystallography,Janssen2018Aperiodic,jiang2014numerical,jiang2024numerical}, introducing the hidden \textit{phason} degree of freedom~\cite{Socolar1986phonons}.
In this framework, phasons correspond to displacements in the perpendicular hyperspace ($\boldsymbol{\beta}_{\perp}$)~\cite{Socolar1986phonons,Janssen2018Aperiodic}, appearing either as spatially varying \textit{phason fluctuations} $\boldsymbol{\beta}_{\perp}(\mathbf r)$ or as a uniform \textit{phason shift} with $\boldsymbol{\beta}_{\perp}(\mathbf r)\equiv \mathbf{C}$.
Phason fluctuations enable low-energy local rearrangements that can assist QC stabilization and growth~\cite{Tang1990Random,Kiselev2012Confirmation,Boissieu2012Phonons,Han2021formation,Je2021entropic,Nagao2015experimental,Wang2025Defect,Schmiedeberg2017Dislocation,Yamada2016atomic}.
By contrast, a uniform phason shift collectively reshuffles atomic motifs, thereby altering the real-space symmetry of QCs while leaving both diffraction patterns and bulk free energy invariant~\cite{Lifshitz2011Symmetry,Widom2008phasons}.
In the absence of any thermodynamic bias, how does nucleation kinetics break this phason-shift degeneracy and select a specific real-space symmetry variant?

Unraveling this question hinges on  characterizing the \textit{critical nucleus (CN)}—the transition state dictating the nucleation barrier and pathway selection. However, nucleation in QCs presents a fundamental puzzle, primarily due to the absence of a periodic template to guide the initial assembly. This inherent difficulty is epitomized by icosahedral quasicrystals (IQCs), which are quasiperiodic in all three dimensions, making them the archetypal system to investigate this problem. 
Experimentally, direct observation of the CN remains elusive due to its transient nature and extreme rarity. Consequently, existing characterizations have focused primarily on the atomic structures of stable IQCs\cite{Takakura2007Atomic,Yamada2016atomic}. In theoretical work, existing simulations have centered on stability and self-assembly\cite{Bedolla2025Relationship,Engel2015Computational,Noya2021How,Pinto2025Automating,Noya2025One,Liang2022Molecular,Baek2025Quasicrystal,Kai2020Stability,Subramanian2016Three,Jiang2017Stability}, leaving the precise calculation of 
CNs and minimum energy paths (MEPs, the most probable nucleation pathways~\cite{Weinan2002String}) unaddressed.
While multi-stage phase transitions and CNs have been identified in 2D QCs~\cite{Yin2021transition,Zhou2024nucleation}, the CNs of 3D IQCs remain uncharted. Most crucially, the intrinsic relationship of nucleation with phason shifts—which govern symmetry—has largely been overlooked.

Here, we address this gap by combining a Landau free-energy model~\cite{Jiang2017Stability} with the spring pair method~\cite{cui2024Spring} to compute the CNs and MEPs for nucleation into both a reference body-centered cubic (BCC) crystal and phason-shifted IQCs.
Unlike the trivial translational degeneracy of periodic crystals, global phason shifts enable multiple symmetry-distinct nucleation pathways.
Specifically, we identify two classes of IQC nucleation pathways distinguished by the symmetry of the CN.
We further show that temperature governs the selection of nucleation pathways by altering their respective energy barriers.
At low temperatures, nucleation proceeds via a direct symmetry-preserving pathway to the ideal IQC (id-IQC) with full icosahedral symmetry.
At higher temperatures, nucleation follows a symmetry-detour pathway that passes through a symmetry-broken CN and yields a reduced-symmetry non-ideal IQC (nid-IQC).
Together, these results establish phasons as the structural origin of nucleation pathway diversity and provide a new physical picture for the emergence of quasiperiodic order from the liquid.

\section{Phason  and Symmetry}
This section explores how phason shifts generate thermodynamically degenerate IQC variants. These variants exhibit identical diffraction intensities yet distinct real-space symmetries, a degeneracy that gives rise to the nucleation selection problem.

\subsection{High-Dimensional Projection Framework}
IQCs lack translational periodicity in three-dimensional (3D) physical space but can be elegantly described as irrational slices of a six-dimensional (6D) periodic hypercubic lattice \cite{jiang2014numerical,jiang2024numerical}. In this framework, the density field $\varphi(\mathbf{r})$ of an IQC is 
\begin{equation}
\varphi(\mathbf{r}) = \sum_{\mathbf{h} \in \mathbb{Z}^6} \hat{\varphi}(\mathbf{h}) e^{i (\mathbf{P}\mathbf{h}) \cdot \mathbf{r}}, \quad \mathbf{r} \in \mathbb{R}^3,
\end{equation}
where $\mathbf{h} \in \mathbb{Z}^6$ are the 6D reciprocal lattice vectors, $\hat{\varphi}(\mathbf{h})$ are the Fourier coefficients.
The projection matrix $\mathbf{P} \in \mathbb{R}^{3 \times 6}$ maps the 6D space to the 3D physical space and is given by
\begin{equation}
\mathbf{P}=
\begin{bmatrix}
1 & \frac{\tau}{2} & \frac{\tau}{2} & \frac{1}{2} & 0 & 0 \\[6pt]
0 & \frac{1}{2} & -\frac{1}{2} & \frac{\tau-1}{2} & 1 & 0 \\[6pt]
0 & \frac{\tau-1}{2} & \frac{1-\tau}{2} & -\frac{\tau}{2} & 0 & 1
\end{bmatrix},
\label{eq:P_matrix}
\end{equation}
where $\tau = (1 + \sqrt{5}) / 2$ is the golden ratio.

\subsection{Phason Degrees of Freedom}
The high-dimensional description includes additional degrees of freedom associated with translations in the 6D superspace.
A superspace translation $\boldsymbol{\beta}\in\mathbb R^6$ modifies the IQC density
through phase shifts in the Fourier modes
\begin{equation}
\varphi_{\boldsymbol{\beta}}(\mathbf{r}) = \sum_{\mathbf{h} \in \mathbb{Z}^6} \hat{\varphi}(\mathbf{h}) e^{i \mathbf{h} \cdot \boldsymbol{\beta}} e^{i (\mathbf{P}\mathbf{h}) \cdot \mathbf{r}}.
\end{equation}
The translation decomposes into orthogonal components
$\boldsymbol{\beta}=\boldsymbol{\beta}_{\parallel}+\boldsymbol{\beta}_{\perp}$
through the corresponding high-dimensional projectors $\Pi_{\parallel}$ and $\Pi_{\perp}$. 

Phonon shifts, corresponding to the parallel component
$\boldsymbol{\beta}_{\parallel}$, induce rigid translations in physical space. 
Phason shifts, corresponding to the perpendicular component
$\boldsymbol{\beta}_{\perp}$, are unique to
quasicrystals. They change the real-space decoration through a collective
rearrangement of atomic positions.

\subsubsection{Thermodynamic Invariants}
For an infinite bulk system, the Landau free energy is invariant under any translation $\boldsymbol{\beta}$, whether it corresponds to a phonon or a phason shift.
As derived from the high-dimensional thermodynamic framework, the free energy depends only on closed sums of wavevectors ($\sum \mathbf{h}_i = 0$), causing the phase factors $e^{i (\sum \mathbf{h}_i) \cdot \boldsymbol{\beta}}$ to cancel out exactly. Consequently,
\begin{equation}
E[\varphi_{\boldsymbol{\beta}}] = E[\varphi], \quad |\hat{\varphi}_{\boldsymbol{\beta}}(\mathbf{h})| = |\hat{\varphi}(\mathbf{h})|.
\end{equation}
This implies that all configurations $\varphi_{\boldsymbol{\beta}}$ connected by phason shifts are \textbf{thermodynamically degenerate} and exhibit identical diffraction patterns.

\subsubsection{Structural Variants}

Thermodynamic degeneracy does not imply identical real-space symmetry. For a given superspace translation $\boldsymbol{\beta}$, the residual rotational symmetry in physical space is controlled solely by the phason shift $\boldsymbol{\beta}_{\perp}$. Consider a global rotation $g\in\mathcal{I}_h$, associated with a 6D unimodular lift $G$. This rotation remains a symmetry of $\varphi_{\boldsymbol{\beta}}$ only when the phason shift satisfies the compatibility condition
\begin{equation}
\Pi_{\perp}\!\big(G^T \boldsymbol{\beta}_{\perp}-\boldsymbol{\beta}_{\perp}\big)\in
\Pi_{\perp}\!\big(2\pi\mathbb{Z}^6\big).
\end{equation}
We therefore define the residual symmetry subgroup as
\begin{equation}
H(\boldsymbol{\beta}_{\perp})
=\Big\{ g\in\mathcal{I}_h \,\Big|\,
\Pi_{\perp}\!\big(G^T \boldsymbol{\beta}_{\perp}-\boldsymbol{\beta}_{\perp}\big)\in
\Pi_{\perp}\!\big(2\pi\mathbb{Z}^6\big) \Big\}.
\end{equation}

When $\boldsymbol{\beta}_{\perp}=0$, the full icosahedral group is preserved and
$H(\boldsymbol{\beta}_{\perp})=\mathcal{I}_h$, which defines the id-IQC. For
generic $\boldsymbol{\beta}_{\perp}\neq 0$, the symmetry reduces to a proper
subgroup of $\mathcal{I}_h$, such as $C_2$, which defines nid-IQC variants.

\begin{figure}[t]
\centering
\includegraphics[width=0.66\columnwidth]{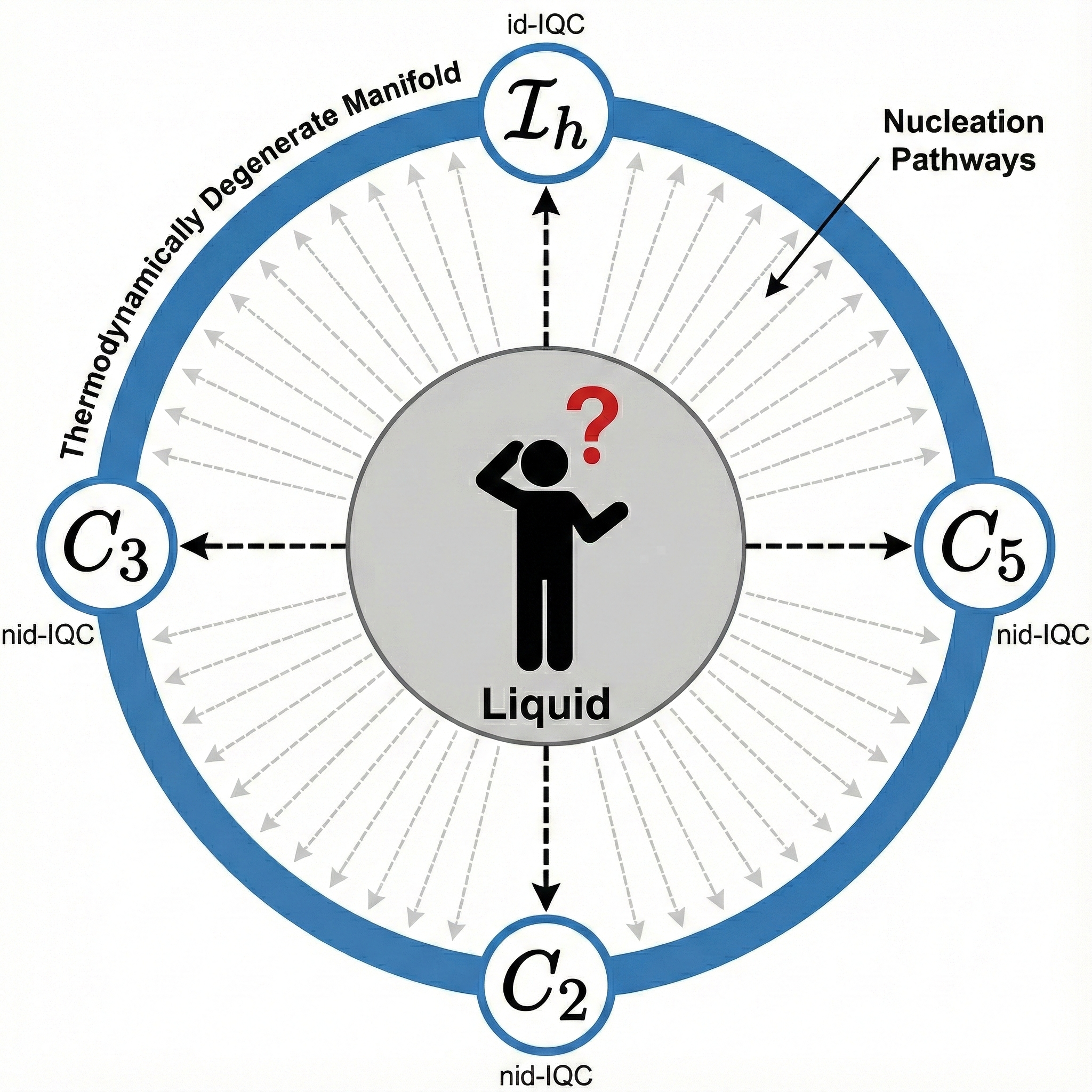}
\caption{
\textbf{Schematic of the nucleation selection dilemma.}
The liquid (center) is surrounded by a thermodynamically degenerate manifold
generated by phason shifts. While these bulk states share the same free energy
and diffraction intensities, they can differ in real-space symmetry,
illustrated by the ideal id-IQC ($\mathcal{I}_h$) and representative nid-IQCs
($C_5$, $C_3$, $C_2$). Dashed arrows indicate competing kinetic routes from the
liquid to different symmetry endpoints. The question mark highlights that bulk
degeneracy does not determine which symmetry is selected during nucleation.
}

\label{fig:nucleation-paradox}
\end{figure}

\subsection{The IQC Nucleation Puzzle}
This framework reveals a kinetic selection problem fundamentally absent in periodic crystals, as summarized schematically in Fig.~\ref{fig:nucleation-paradox}.
Global phason shifts generate a continuous family of IQC variants that are
thermodynamically degenerate yet structurally distinct in real space.

The thermodynamic degeneracy of these bulk variants implies that thermodynamics alone cannot dictate the selection of a structural variant during nucleation. Instead, phason-induced structural diversity implies that nucleation may initiate from distinct local motifs, resulting in CNs with disparate symmetries. Consequently, we investigate whether these structural differences translate into distinct nucleation barriers, thereby providing a mechanism for kinetic selection. By computing the CNs and their corresponding MEPs, we quantify this selection process across representative IQC variants.

% \hlred{
% This phason-enabled structural multiplicity therefore
% suggests competing, symmetry-distinct nucleation routes.
% To quantify pathway diversity, we compute the CNs and
% the associated MEPs for IQC variants.}

% Because these variants are degenerate in bulk free energy, thermodynamics alone
% does not determine which symmetry emerges upon nucleation from an undercooled
% liquid. 
% \hlcy{This degeneracy therefore shifts the selection problem to kinetics: whether phason shifts yield symmetry-distinct CNs and nucleation pathways, what kinetic advantage selects the dominant pathway, and how the nucleation-and-growth mechanism varies across pathways.}

\section{Model}
The system is described by a Landau free-energy functional of a 
density field $\varphi(\mathbf r)$ on a domain $\Omega$, where $\varphi(\mathbf r)$
denotes the deviation from a reference liquid state, and satisfies the
mass-conservation constraint $\int_{\Omega}\varphi(\mathbf r)\,d\mathbf r = 0$.
Within this framework, the free energy is given by
\begin{equation}\label{formula1}
\begin{aligned}
E[\varphi]
&=\frac{1}{2}\iint_{\Omega\times\Omega}
\varphi(\mathbf r)\,G(|\mathbf r-\mathbf r'|)\,\varphi(\mathbf r')\,
d\mathbf r\,d\mathbf r' \\
&\quad + \int_{\Omega}\left(
\frac{\varepsilon}{2}\varphi^2(\mathbf r)
-\frac{\alpha}{3}\varphi^3(\mathbf r)
+\frac{1}{4}\varphi^4(\mathbf r)
\right)\,d\mathbf r ,
\end{aligned}
\end{equation}
where $\varepsilon$ is a temperature-like control parameter that penalizes density
fluctuations and $\alpha$ characterizes the intensity of three-body interaction \cite{Jiang2015Stability,Barkan2011Stability}.
In this work, we study nucleation in a parameter regime where the homogeneous
liquid $\varphi(\mathbf r) = 0$ is a local minimum of $E[\varphi]$, while the ordered
phases of interest correspond to deeper minima.

The system is governed by an effective pair interaction potential of the
Gaussian-polynomial form,
\begin{equation}
G(r)=\exp\!\left(-\frac{\sigma^2 r^2}{2}\right)
\left(c_0+c_2 r^2+c_4 r^4+c_6 r^6+c_8 r^8\right),
\label{eq:interaction_potential}
\end{equation}
which has been widely used to model crystallization and quasicrystal formation
\cite{Ratliff2019Which,Jiang2017Stability,Barkan2014Controlled}.

\begin{figure}[!hbpt]
\centering
\includegraphics[width=1.0\columnwidth]{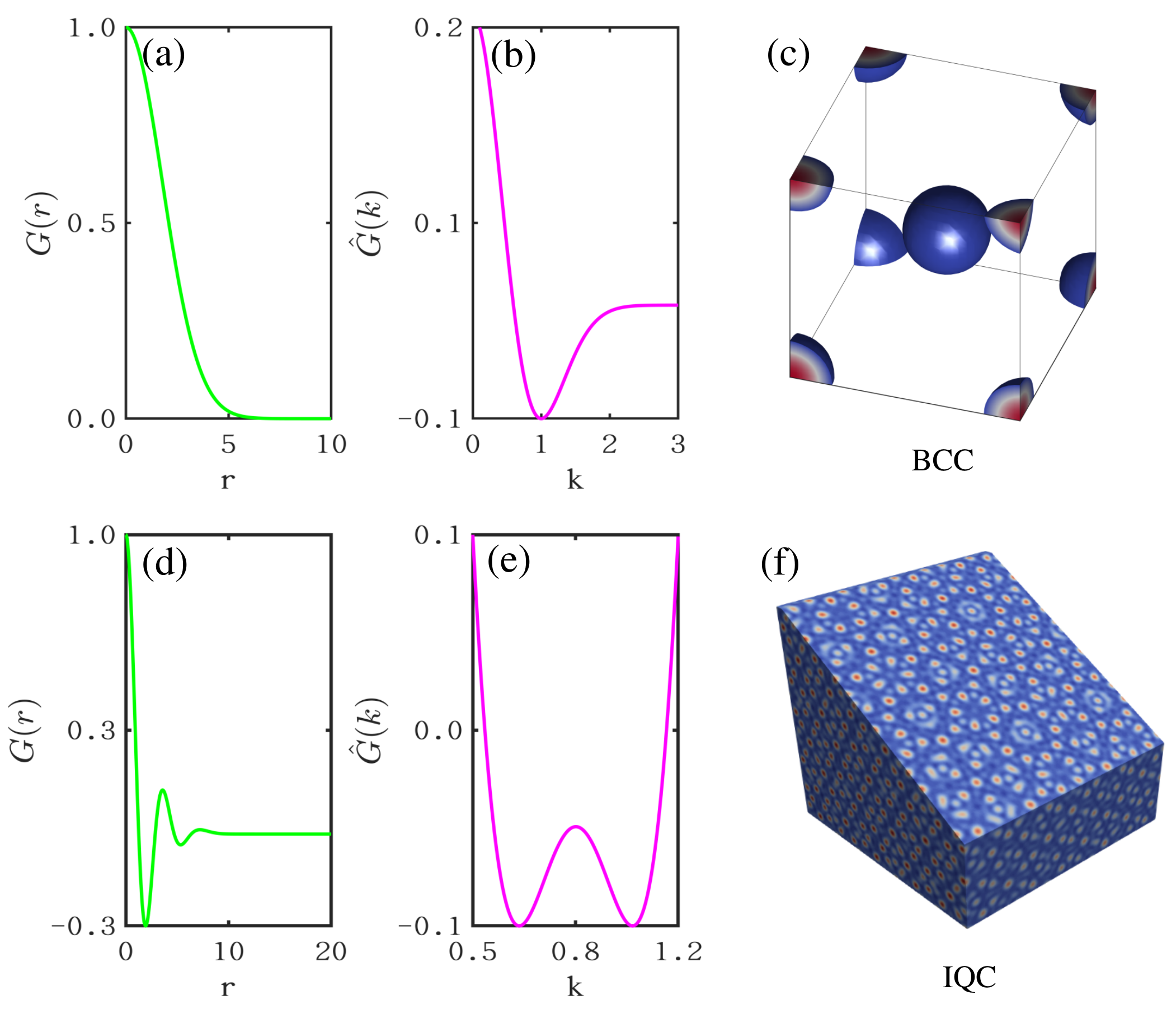}~~~
\caption{\textbf{Pair interaction potentials and the ordered structures they stabilize in the present study.}
(a) Real-space pair interaction potential $G(r)$ that is purely repulsive.
(b) Fourier transform $\hat{G}(k)$ exhibiting a single dominant characteristic length scale.
(c) The body-centered cubic (BCC) crystal stabilized by this interaction.
(d) Real-space pair interaction potential $G(r)$ with alternating attractive and repulsive components (sign-changing).
(e) Fourier transform $\hat{G}(k)$ exhibiting two characteristic length scales whose ratio equals the golden ratio.
(f) The icosahedral quasicrystal (IQC) stabilized by this interaction.}
\label{interaction_potential}
\end{figure}

By tuning the parameters in Eq.~(\ref{eq:interaction_potential}), the interaction potential stabilizes either periodic crystals with a single characteristic length scale (exemplified by the BCC phase) or IQCs involving two length scales related by the golden ratio (Fig.~\ref{interaction_potential}). The specific parameter sets employed in this work are carefully chosen to stabilize these distinct ordered phases.

\section{Methods}

Within the Landau free-energy framework, the undercooled liquid and ordered
phases correspond to local minima of the free-energy functional $E[\varphi]$.
The CN is an index-1 saddle point that separates the liquid basin from that of a given ordered phase. For each nucleation route, we
characterize nucleation and subsequent growth along the associated MEP on $E[\varphi]$, and define the nucleation barrier as
$\Delta E = E_{\mathrm{CN}} - E_{\mathrm{liq}}$.

\subsection{Periodic Approximation Method and  Discretization}

To accurately simulate the nucleation of aperiodic QCs within a finite computational domain, we employ the Periodic Approximation Method (PAM) \cite{ Jiang2025On} to reduce boundary artifacts. This approach has been successfully validated for two-dimensional QC systems \cite{Yin2021transition, Zhou2024nucleation}. Specifically, we approximate the 
IQC density field $\varphi(\mathbf{r})$ using a periodic approximant defined on a cubic domain $\Omega = [0, 2\pi L]^3$, where the integer $L$ determines the domain size.

The approximation proceeds in two stages. First, to enforce periodicity,
the projected reciprocal vectors $\boldsymbol{\lambda}=\mathbf{P}\mathbf{h}\in\mathbb{R}^3$
(which typically have irrational components) are approximated by rational vectors
$\mathbf{k}/L\in\mathbb{Q}^3$, yielding a periodic approximant

\begin{equation}
\label{eq:periodic_approx_step1}
\varphi(\mathbf{r}) = \sum_{\mathbf{h} \in \mathbb{Z}^6} \hat{\varphi}(\mathbf{h}) e^{i \mathbf{P}\mathbf{h} \cdot \mathbf{r}} \approx \sum_{\mathbf{k} \in \mathbb{Z}^3} \hat{\varphi}(\mathbf{k}) e^{i \frac{\mathbf{k}}{L} \cdot \mathbf{r}},
\end{equation}

\noindent where $\mathbf{k}=\{\lfloor \lambda_j L\rceil\}_{j=1}^3$, with $\lfloor x\rceil$ denoting rounding to the nearest integer. For the IQC, the coefficients to be approximated simultaneously—derived from the elements of the projection matrix (Eq.~\ref{eq:P_matrix})—are $\frac{1}{2}$ and $\frac{\tau}{2}$. The resulting approximation error is quantified by the Diophantine error \cite{Davenport1946Simultaneous}

\begin{equation}
\label{eq:EDA}
E_{DA}(L)=\left\| \left(\frac{1}{2}, \frac{\tau}{2} \right) - \frac{1}{L}\left( \left\lfloor \frac{L}{2} \right\rceil ,  \left\lfloor \frac{\tau L}{2} \right\rceil  \right)
\right\|_\infty.
\end{equation}

The local minima of $E_{DA}$ occur at $L=16, 26, 42, \dots$.
Numerical verifications confirm that these minima with $L \ge 26$ ensure  both sufficient numerical accuracy and a domain volume large enough to contain the CN.
Consequently, we set $L=26$ for all IQC simulations, yielding $E_{DA}(26) \approx 1.32 \times 10^{-3}$.

Second, for numerical implementation, we truncate the Fourier spectrum,
\begin{equation}
\label{eq:spectral_set}
\mathbf{K}_N^3=\left\{\mathbf{k}\in\mathbb{Z}^3:\;
-\frac{N}{2}\le k_j<\frac{N}{2},\; j=1,2,3 \right\},
\end{equation}
\noindent which yields the truncated expansion
\begin{equation}
\label{eq:periodic_approx_step2}
\varphi(\mathbf{r})\approx
\sum_{\mathbf{k}\in\mathbf{K}_N^3}\hat{\varphi}(\mathbf{k})\, e^{i\frac{\mathbf{k}}{L}\cdot\mathbf{r}}.
\end{equation}
\noindent We discretize Eq.~\ref{eq:periodic_approx_step2} using a Fourier
pseudo-spectral method with $N=256$ grid points in each dimension, enabling
efficient evaluation of nonlinear convolution terms via the fast Fourier transform.

\subsection{Spring Pair Method}
%\label{subsec:SPM}
We identify CNs by locating index-1 saddle points of the discretized
free-energy functional $E[\varphi]$ using the SPM~\cite{cui2024Spring}.
SPM is a gradient-only saddle-search scheme that evolves two spring-coupled
replicas and converges to a saddle without Hessian evaluations, while providing
an estimate of the unstable mode.

For each parameter set, we initialize SPM from the homogeneous liquid state
(with small mean-zero random perturbations) and repeat the search from multiple
independent initializations to sample distinct CNs.
Given a converged saddle $\varphi^{\ast}$ and its unstable mode $\mathbf{u}$, we
obtain the associated MEP by relaxing
$\varphi^{\ast}\pm\delta\,\mathbf{u}$ via gradient descent. The two downhill
branches connect the liquid basin and the target ordered basin~\cite{cui2024Spring}.

\subsection{Structural Analysis and Visualization}
Particle positions are extracted as the coordinates of local maxima of the
density field $\varphi(\mathbf r)$. Structural visualization and rendering are
performed using OVITO.

\section{Results and Discussion}

\begin{figure*}[t]
\centering
\includegraphics[width=0.666\textwidth]{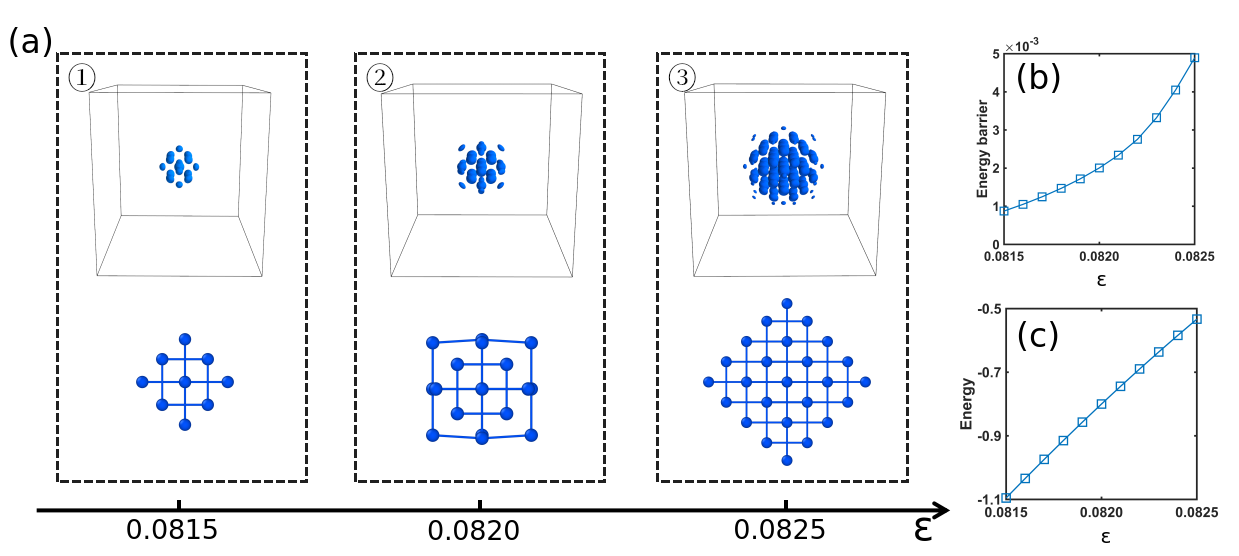}\\
\includegraphics[width=0.666\textwidth]{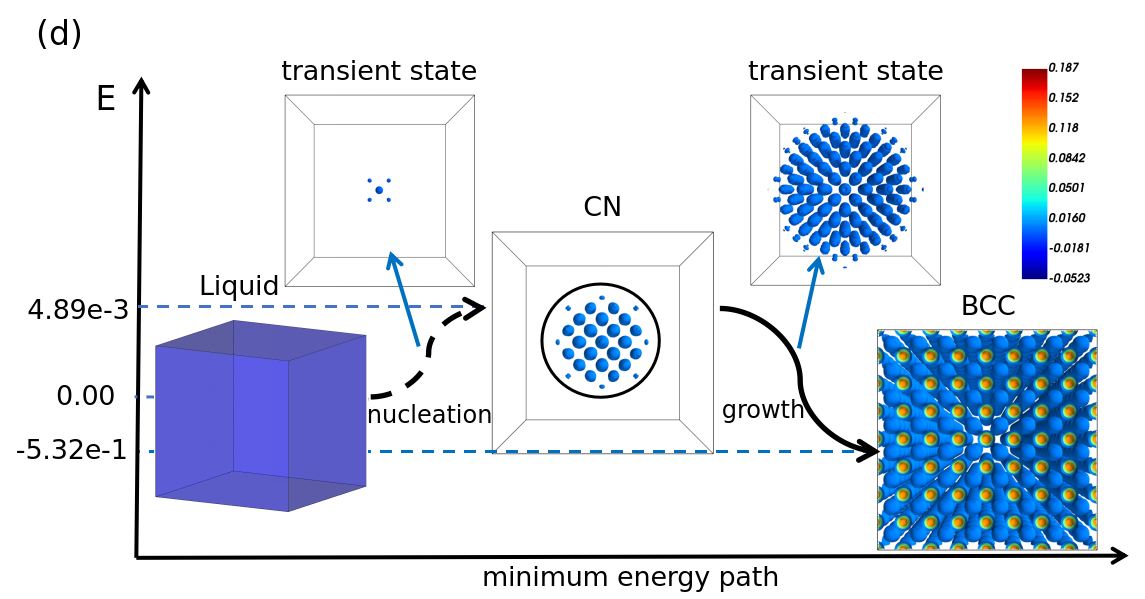}
\caption{\textbf{Nucleation of the BCC crystal as a reference case.}
(a) Evolution of BCC CNs at increasing temperatures
($\varepsilon = 0.0015$, $0.0020$, and $0.0025$; $\alpha = 0.2$). Upper panels show
density distributions, and lower panels show the corresponding atomic
arrangements with BCC connectivity. The CNs preserve BCC symmetry while increasing in size.
(b) Nucleation energy barrier as a function of temperature.
(c) Free energy of the bulk BCC phase grown from the CNs, indicating
a weakening thermodynamic driving force with increasing temperature.
(d) MEP for BCC nucleation and growth at $\varepsilon = 0.0025, \alpha = 0.2 $.}
\label{FIG-BCC-critical-nuclei}
\end{figure*}

\subsection{Nucleation of BCC Crystal: A Reference Case}
Before addressing IQC nucleation, we examine BCC crystallization as a reference system.
For this classic periodic crystal, all thermodynamically equivalent configurations differ only by rigid translations in physical space, corresponding to phonon shifts, with no phason degrees of freedom. 
This makes BCC nucleation an ideal baseline for contrasting periodic and quasiperiodic ordering.

Figure~\ref{FIG-BCC-critical-nuclei} shows the temperature dependence of BCC
nucleation. Over the entire range of $\varepsilon$ investigated, the critical
nuclei retain  BCC symmetry, while their size and the associated
nucleation barrier increase monotonically with temperature
[Fig.~\ref{FIG-BCC-critical-nuclei}(a,b)]. Meanwhile, the free energy of the bulk
BCC phase increases with increasing temperature
[Fig.~\ref{FIG-BCC-critical-nuclei}(c)], reflecting a weakening thermodynamic driving force
for crystallization, consistent with classical crystal nucleation
theory. The corresponding MEP at $\varepsilon = 0.0025$ [Fig.~\ref{FIG-BCC-critical-nuclei}(d)], typical of the entire range investigated, reveals a direct, symmetry-preserving pathway characterized by the simple accumulation of BCC unit cells.

Since the degeneracy in BCC crystals stems solely from translational symmetry, all nucleation pathways leading to thermodynamically equivalent BCC states are identical up to a rigid translation. Even
if nonclassical mechanisms such as two-step nucleation~\cite{DeYoreo2013More,James2019How,Li2021Temp}
occur, this translational equivalence remains unchanged.
% This behavior contrasts with IQC nucleation,  where phason degrees of freedom enable multiple nucleation pathways with
% different critical nucleus symmetries.

\subsection{Nucleation and Growth of IQCs}

\begin{figure*}[!hbpt]
\centering
\includegraphics[width=1.618\columnwidth]{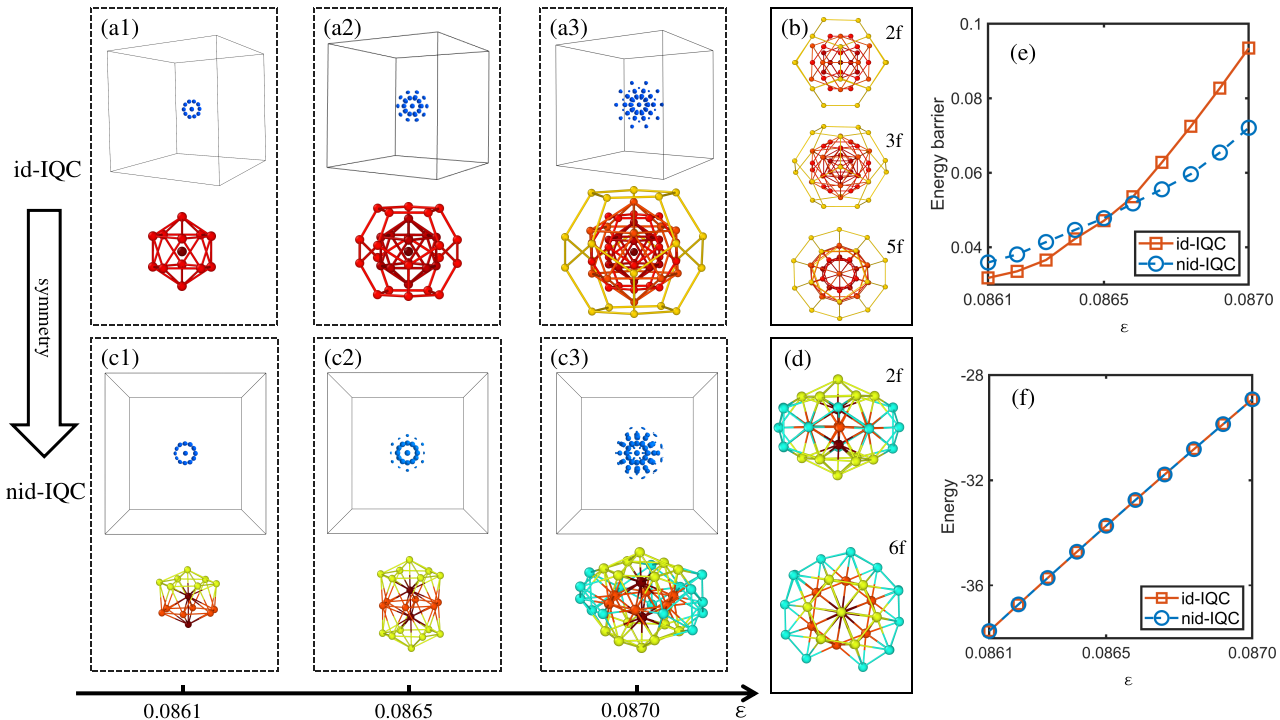}
\caption{\textbf{Temperature-dependent CNs for id-IQC and nid-IQC and the crossover in nucleation barriers.}
(a1--a3) Density distributions (upper) and corresponding atomic arrangements (lower) of id-IQC CNs at
$\varepsilon=0.0861$, $0.0865$, and $0.0070$ ($\alpha=0.3$), showing robust $\mathcal{I}_h$ symmetry.
(b) Atomic arrangement of (a3) viewed along 2-fold, 3-fold, and 5-fold axes, confirming full icosahedral symmetry.
(c1--c3) CNs along the nid-IQC pathway at the same temperatures, exhibiting reduced symmetry with
approximate 6-fold motifs.
(d) Atomic arrangement of (c3) viewed along a 2-fold axis and an approximate 6-fold axis, highlighting symmetry reduction.
(e) Nucleation barriers for id-IQC and nid-IQC versus $\varepsilon$, showing a crossover near $\varepsilon\approx0.0865$.
(f) Energies of the fully developed id-IQC and nid-IQC states versus $\varepsilon$, indicating degenerate
thermodynamic stability.}
\label{FIG-IQC-Critical-Nuclei}
\end{figure*}

\begin{figure}[!hbpt]
\centering
\includegraphics[width=0.7\columnwidth]{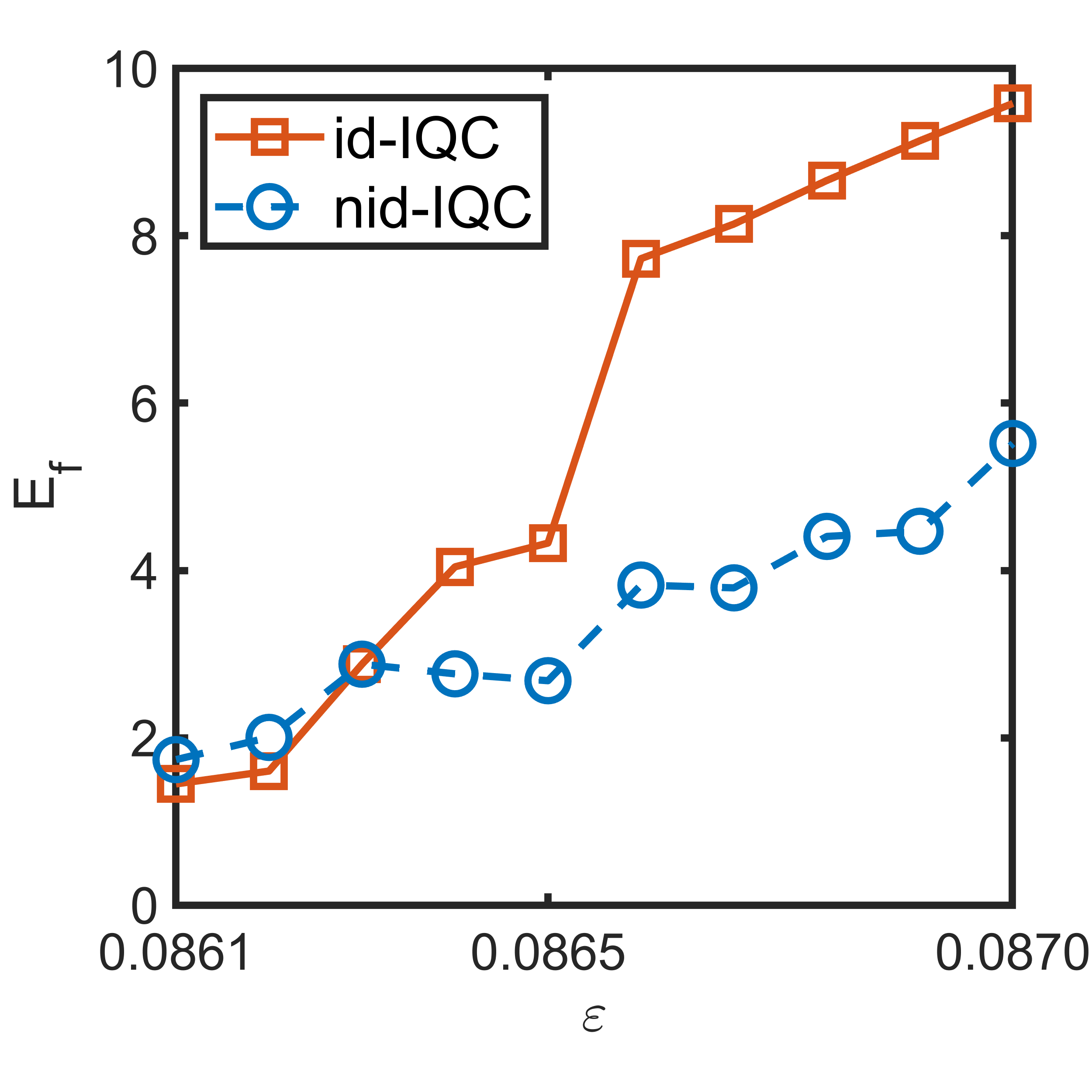}~~~
\caption{Energy penalty for density fluctuations ($E_f = \int_{\Omega} \frac{\varepsilon}{2}\varphi^2 dr$) in CNs as a function of temperature parameter $\varepsilon$ for id-IQC (solid line with squares) and nid-IQC (dashed line with circles). At lower temperatures, both types of CNs exhibit similar energy penalties. However, as temperature increases, the energy penalty for id-IQC CNs rises more rapidly than for nid-IQC CNs, creating a significant energy advantage for the lower-symmetry nid-IQC nucleation pathway at higher temperatures. This demonstrates why systems preferentially form CNs with lower symmetry at elevated temperatures.}\label{fig:Energy_penalty}
\end{figure}

\subsubsection{Two Distinct IQC CNs}

Unlike the trivial translational degeneracy in periodic crystals, phason degrees of freedom render IQCs thermodynamically degenerate yet structurally distinct. 
This phason-enabled structural multiplicity may give rise to pathway diversity via symmetry-distinct CNs.

To quantify this, we computed the CNs and MEPs for nucleation from the supercooled liquid into both the id-IQC and nid-IQC. Figure~\ref{FIG-IQC-Critical-Nuclei} presents two representative pathways that reveal two symmetry-distinct classes of IQC nucleation, based on the symmetry of the CN. 

In the \emph{id-IQC nucleation pathway}, the CN consistently exhibits full
icosahedral symmetry ($\mathcal{I}_h$) over a wide temperature range
[Fig.~\ref{FIG-IQC-Critical-Nuclei}(a,b)]. The atomic configuration in
Fig.~\ref{FIG-IQC-Critical-Nuclei}(b) confirms that the CN shares the symmetry
and key structural motifs of the final id-IQC steady state. We refer to such
paths as \emph{symmetry-preserving} routes.

In contrast, the \emph{nid-IQC nucleation pathway} proceeds through a
reduced-symmetry CN. For nucleation toward a representative $C_2$-symmetric
nid-IQC, the CN displays pronounced pseudo-sixfold local motifs rather than
icosahedral order [Fig.~\ref{FIG-IQC-Critical-Nuclei}(c,d)], indicating a
\emph{symmetry-detour} route in which the CN is symmetry-mismatched with the final
state.

Temperature determines pathway selection by modifying their nucleation barriers. Figure~\ref{FIG-IQC-Critical-Nuclei}(e) shows a clear temperature-dependent crossover. At lower temperatures, the symmetry-preserving id-IQC pathway faces a lower barrier and is kinetically favored. However, as temperature increases, the id-IQC barrier rises steeply, leading to a crossover near $\varepsilon \approx 0.0865$. Beyond this point, the symmetry-detour nid-IQC pathway becomes the preferred kinetic route. Crucially, since the free energies of the developed id-IQC and nid-IQC bulk phases are identical [Fig.~\ref{FIG-IQC-Critical-Nuclei}(f)], this selection is purely kinetic rather than thermodynamic.

The physical origin of this crossover lies in the competition between the energetic gain from ordering and the quadratic penalty for density modulations. At lower temperatures (small $\varepsilon$), the quadratic cost $E_f=\int_{\Omega}\frac{\varepsilon}{2}\varphi^2\,d\mathbf r$ is small, allowing the formation of a fully $\mathcal{I}_h$-symmetric nucleus with modest penalty. At higher temperatures, however, enforcing perfect $\mathcal{I}_h$ symmetry necessitates high-amplitude density modulations to maintain structural order, causing $E_f$ to rise sharply. By temporarily relaxing the symmetry constraint, the nid-IQC nucleus avoids this steep penalty, forming with weaker modulations and a lower $E_f$. As quantified in Fig.~\ref{fig:Energy_penalty}, this differential sensitivity drives the preference for the lower-symmetry pathway at elevated temperatures.

In summary, these results demonstrate that phason degrees of freedom unlock pathway diversity unavailable to periodic crystals. Temperature selects the kinetically favored route by tuning the cost of symmetry.

\subsubsection{Nucleation and Growth Process of id-IQC}

\begin{figure*}[!hbpt]
\centering
\includegraphics[width=1.6\columnwidth]{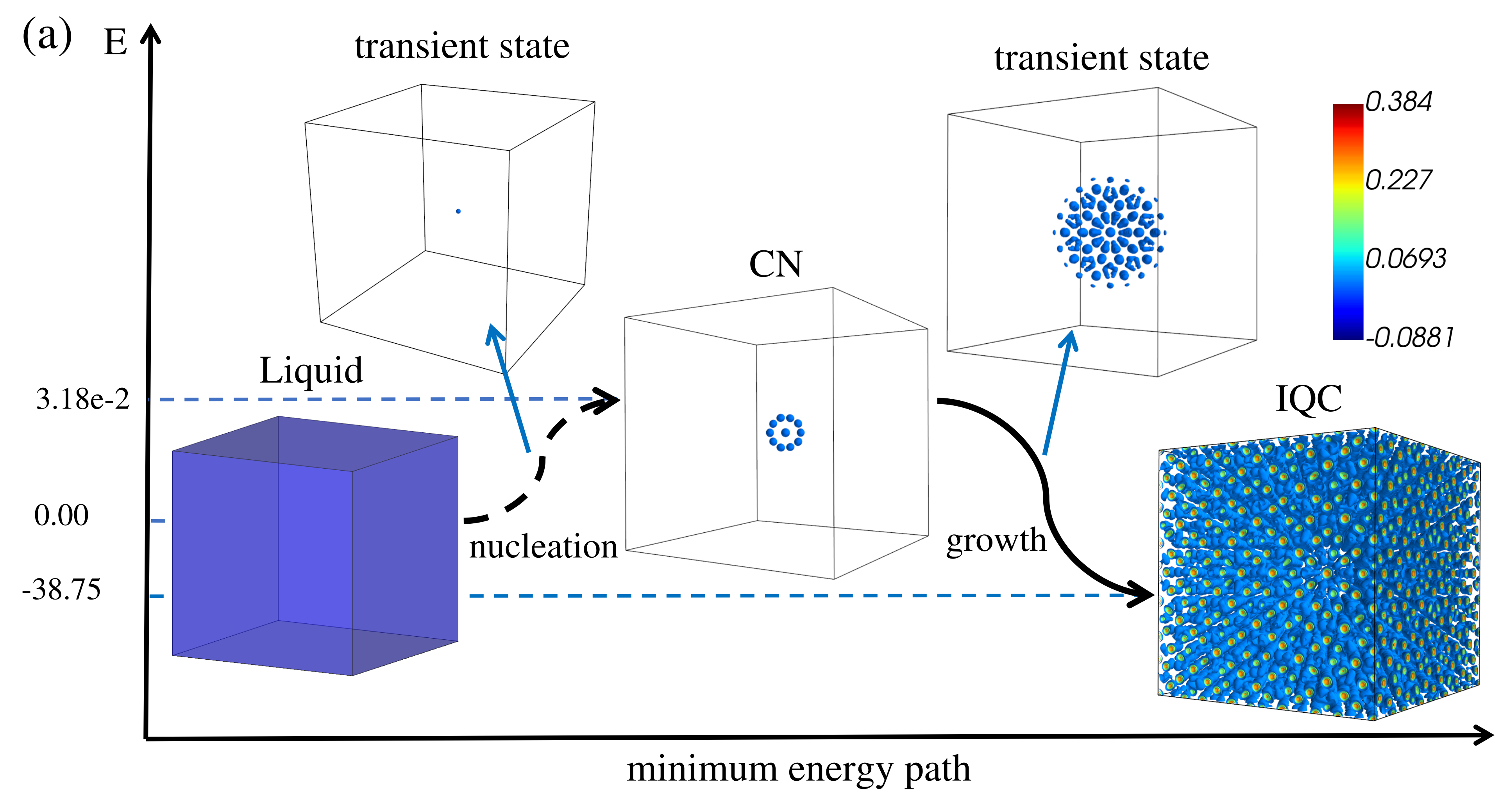}\\
\includegraphics[width=1.5\columnwidth]{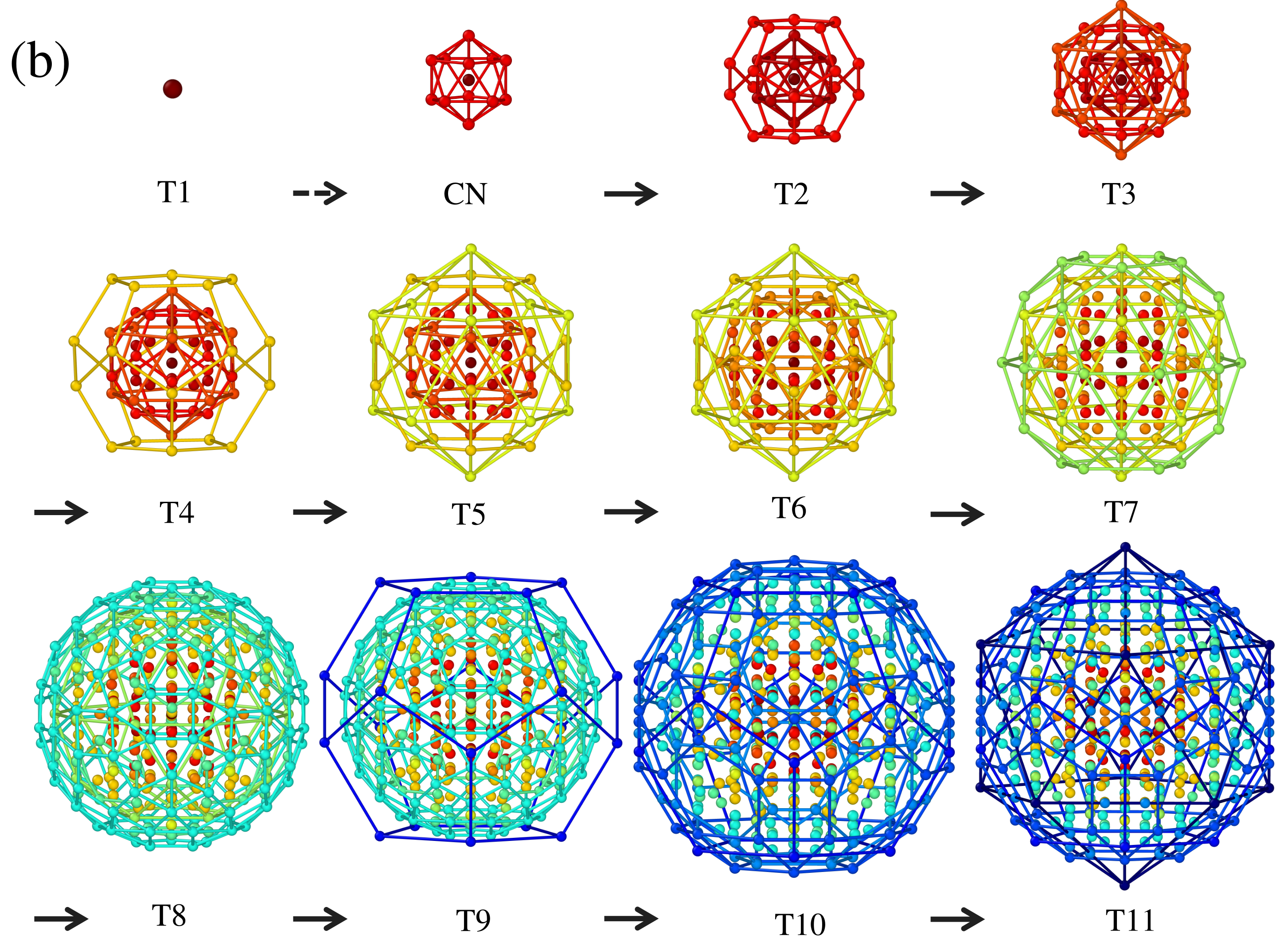}
\caption{\textbf{Nucleation and growth process of id-IQC along the MEP at
$\varepsilon = 0.0861$ and $\alpha = 0.3$.}
(a) MEP from the supercooled liquid (DIS) to the stable id-IQC.
(b) Atomic configurations sampled along the MEP, illustrating nucleation
(T1--CN) and growth (CN--T11). Particles are placed at local density maxima, 
with connections shown only for the three outermost layers for clarity. Colors encode
the radial distance from the nucleus center (inner to outer). Nucleation starts
from a small seed (T1) and reaches a fully developed icosahedral CN. Beyond CN, growth follows the icosahedral--dodecahedral dual motif
with alternating shells, while additional $\mathcal{I}_h$-symmetric polyhedral motifs
(at T7, T8, and T10) appear
between neighboring icosahedral and dodecahedral shells to fill local voids and
improve packing.}
\label{FIG-id-MEP}
\end{figure*}

Having established the kinetic selection within these two pathway classes, we now examine their distinct microscopic nucleation and growth mechanisms by analyzing the structural evolution along the MEPs.

We first focus on the \emph{symmetry-preserving id-IQC route}. Figure~\ref{FIG-id-MEP}(a)
shows a representative MEP at $\varepsilon=0.0861$, connecting the
disordered liquid directly to the id-IQC steady state. Structural evolution
along this path reveals a highly ordered, hierarchical assembly. As shown in
Fig.~\ref{FIG-id-MEP}(b), nucleation initiates from a compact seed (T1) and rapidly
develops into a CN that exhibits full $\mathcal{I}_h$
symmetry and is structurally commensurate with the final id-IQC.

We term the subsequent growth mode ``duality growth'', a mechanism that
exploits the geometric duality between icosahedra and dodecahedra to maintain
symmetry-preserving growth. Specifically, dodecahedral vertices emerge along the
outward normals of icosahedral faces, while new icosahedral vertices form along
the normals of dodecahedral faces. This alternation preserves full
$\mathcal{I}_h$ symmetry while extending quasiperiodic order outward.

Beyond the primary shells, the growing nucleus incorporates auxiliary
$\mathcal{I}_h$-symmetric motifs (e.g., 32-faced polyhedra and
rhombicosidodecahedra at T7, T8, and T10) to fill local voids between shells,
thereby improving packing and enabling the continuous extension of quasiperiodic
order to the macroscopic scale.

\subsubsection{Nucleation and Growth Process of nid-IQC}

\begin{figure*}[!hbpt]
\centering
\includegraphics[width=0.8\textwidth]{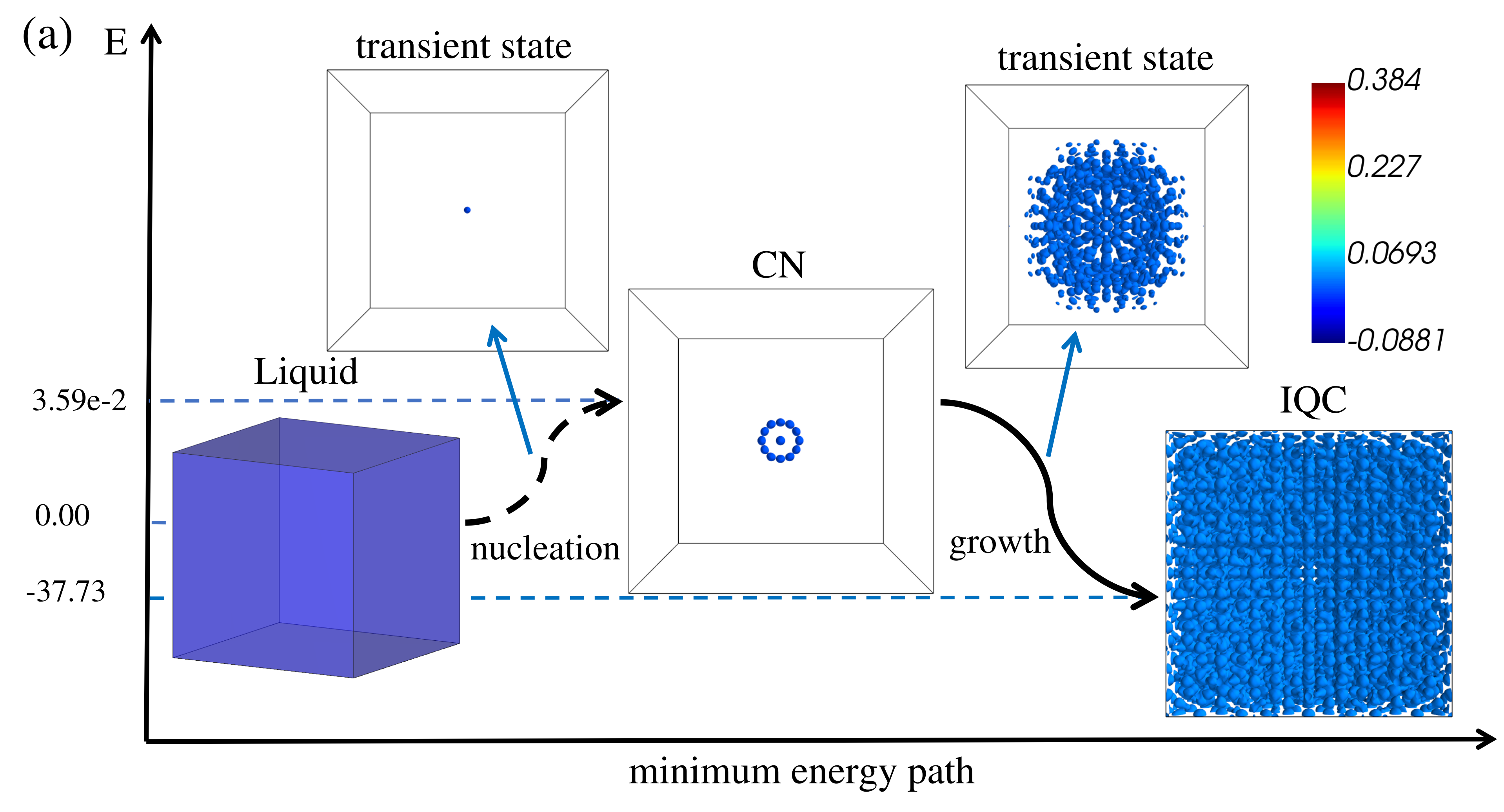}
\includegraphics[width=0.75\textwidth]{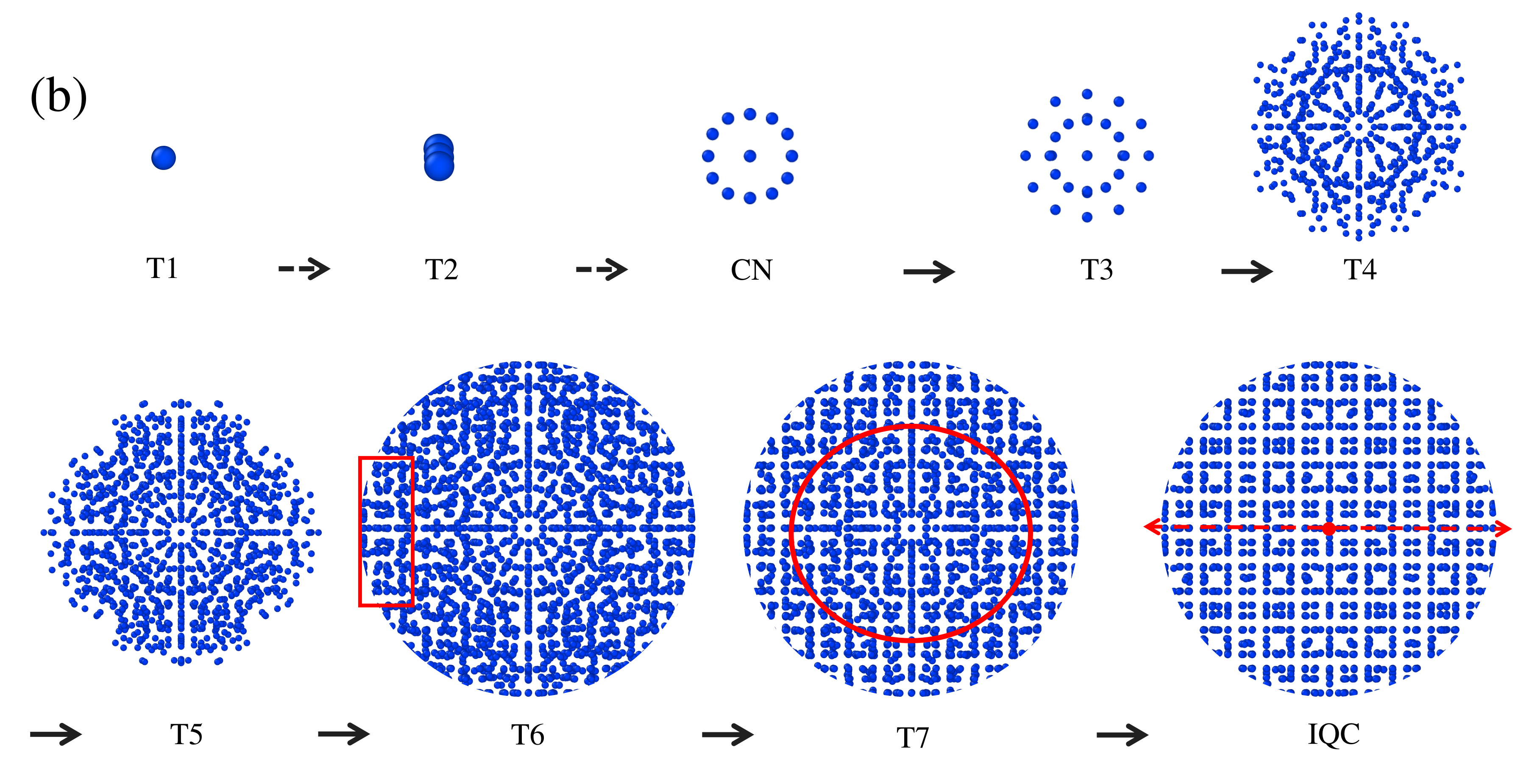}
\caption{\textbf{Nucleation and growth process of a representative nid-IQC along the MEP at $\varepsilon = 0.0861$ and $\alpha = 0.3$.}
(a) Density profiles sampled along the MEP from the DIS to the nid-IQC, with the CN indicated.
(b) Atomic configurations sampled along the MEP. The global symmetry follows a detour: DIS $\rightarrow$ a CN with pronounced sixfold ($C_6$) features $\rightarrow$ the $C_2$-symmetric nid-IQC steady state. From CN to T6, the $C_6$-symmetric nucleus acts as a temporary scaffold. Growth then proceeds via the progressive formation of characteristic local IQC clusters around this core (active regions marked by red rectangles). In the final stage (T6--IQC), atoms within the core rearrange to eliminate the symmetry mismatch, yielding the final $C_2$-symmetric IQC.}
\label{FIG-nid-MEP}
\end{figure*}

In stark contrast to the symmetry-preserving id-IQC route, Figure~\ref{FIG-nid-MEP} details a representative \emph{symmetry-detour pathway} leading to a $C_2$-symmetric nid-IQC steady state (shown at $\varepsilon=0.0861$). Unlike the id-IQC pathway (Figure~\ref{FIG-id-MEP}), where the CN is symmetry-consistent with the final state, the nid-IQC CN is \emph{symmetry-mismatched}. It exhibits a pronounced $C_6$-symmetric character that diverges sharply from the $C_2$ symmetry of the eventual steady state.

Structural snapshots in Figure~\ref{FIG-nid-MEP}(b) reveal a ``transient scaffolding'' mechanism. The nucleation event establishes a $C_6$-symmetric core that functions not as a permanent template, but as a temporary structural scaffold. During the intermediate growth stage (CN$\rightarrow$T6), this scaffold expands and consolidates, while characteristic local IQC clusters progressively form around it (active regions highlighted at T6), resulting in a heterogeneous intermediate structure.

Crucially, the final symmetry emerges only at the late stages of growth. From T7 to the final $C_2$ nid-IQC state, atoms within the central $C_6$-symmetric scaffold undergo a collective rearrangement that eliminates the symmetry mismatch with the surrounding shell, producing a defect-free, globally $C_2$-symmetric structure. 
Overall, unlike classical nucleation in which the CN templates the final symmetry, the symmetry-detour pathway is scaffolded by a $C_6$-symmetric core that dynamically reorganizes to yield the final nid-IQC state.

\section{Discussion}

\begin{figure}[!htbp]
\centering
\includegraphics[width=1.0\columnwidth]{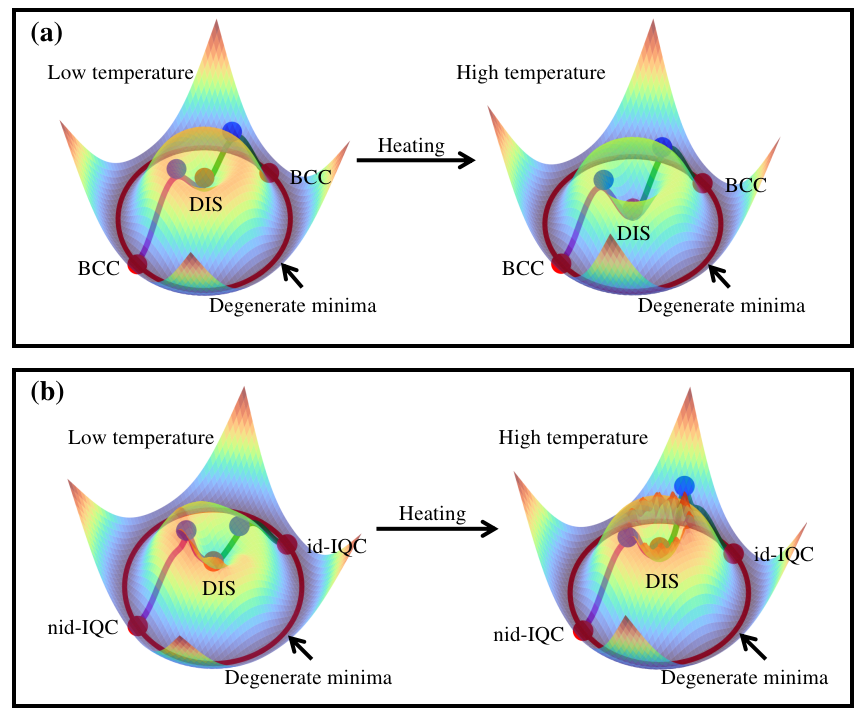}
\caption{\textbf{Schematic energy landscapes for BCC and IQC nucleation.}
(a) BCC. Red circles denote degenerate BCC minima related by rigid
translations in physical space. Blue dots denote the saddle points (CN). Representative nucleation pathways (green and purple) are translation-equivalent
and therefore have identical barrier heights. Increasing temperature primarily
raises the barrier without changing the pathway class.
(b) IQC. Red circles denote thermodynamically degenerate IQC minima parameterized
by higher-dimensional translations $\boldsymbol{\beta}= \boldsymbol{\beta}_\parallel +
\boldsymbol{\beta}_\perp$. Blue dots denote the
corresponding saddle points. Two representative symmetry-distinct routes are illustrated, an
id-IQC route (green) and a nid-IQC route (purple), whose relative barrier
heights depend on temperature.}
\label{fig:energy_landscape}
\end{figure}

By combining a Landau free-energy model with the SPM, we identify nucleation
pathways for IQCs. The essential difference from periodic crystals is the
presence of phason degrees of freedom. Phason shifts generate a degenerate
family of structurally distinct steady states and, already at nucleation, allow
symmetry-distinct CNs with different barriers
(Fig.~\ref{FIG-IQC-Critical-Nuclei}).

For periodic crystals such as BCC, degeneracy arises solely from rigid translation. All translated states share identical symmetry and local coordination. Consequently, nucleation paths are translation-equivalent, and their CNs differ only by spatial position (Fig.~\ref{FIG-BCC-critical-nuclei}). Increasing temperature simply scales the barrier height without introducing new structural choices.

In contrast, IQC degeneracy originates from translations in a higher-dimensional
embedding space. The parallel component $\boldsymbol{\beta}_\parallel$
corresponds to trivial rigid shifts, whereas the perpendicular component
$\boldsymbol{\beta}_\perp$ (phasons) modifies the real-space
symmetry. This generates a family of IQC steady states ranging from the
$\mathcal{I}_h$-symmetric id-IQC to symmetry-broken nid-IQC variants. This phason-driven structural diversity naturally leads to multiple nucleation
routes, yielding CNs with distinct symmetries (Fig.~\ref{FIG-IQC-Critical-Nuclei}).

Temperature selects among these routes by reshaping the relative barrier
heights (Fig.~\ref{FIG-IQC-Critical-Nuclei}(e)). At low temperatures, the direct symmetry-preserving route toward the
id-IQC is favored (Fig.~\ref{FIG-id-MEP}). At higher temperatures, maintaining perfect $\mathcal{I}_h$
symmetry in the CN incurs a steep energetic penalty from the
quadratic cost of large-amplitude density modulations. The system then lowers
the barrier by taking a symmetry detour via a lower-symmetry CN,
leading to the nid-IQC (Fig.~\ref{FIG-nid-MEP}).

Taken together, our findings establish phasons as the structural origin of pathway diversity, providing a new physical picture for the emergence of quasiperiodic order (Fig.~\ref{fig:energy_landscape}).

\nocite{*}
\bibliography{reference}% Produces the bibliography via BibTeX.

\end{document}